\documentclass[a4paper,11pt]{article}
\usepackage{pos}

\title{The stochastic approach for anomalies in supersymmetric theories}
\author*[a]{Stam Nicolis}

\affiliation[a]{Institut Denis Poisson, Université de Tours, Université d'Orléans, CNRS (UMR7013)\\
  Parc Grandmont, 37200 Tours, France}

\emailAdd{stamatios.nicolis@univ-tours.fr}

\abstract{We discuss how the stochastic approach for supersymmetric theories leads to new ways of characterizing anomalies in how supersymmetry can be broken.}

\FullConference{Proceedings of the Corfu Summer Institute 2025 "School and Workshops on Elementary Particle Physics and Gravity" (CORFU2025)\\
27 April - 28 September, 2025\\
Corfu, Greece\\}

\begin{document}
\maketitle

\section{Introduction}\label{intro}
It is well known that global symmetries can be subject to ``anomalies'', in that fluctuations may not preserve all the properties of the classical equations of motion. This provides a ``third way'' for understanding the relevance of symmetries, beyond the ``Wigner mode'' (the symmetry leaves the action and the solutions of the classical equations of motion invariant and, if it is continuous, there are the corresponding conserved quantities) and the Nambu--Goldstone mode (the symmetry leaves the action invariant, but not all solutions of the classical equations of motion and the solutions that are not invariant are, in fact, preferred): The identities, satisfied by the correlation functions, acquire additional contributions.

Rigid supersymmetry is a global symmetry and can, therefore, be subject to anomalies, when fluctuations are relevant. Understanding them is, naturally, useful for understanding how supersymmetry can be realized and broken. There has been considerable work in the subject~\cite{bzowski2020consistency,katsianis2020supersymmetry,Papadimitriou:2017kzw,Kuzenko:2019vvi} (of course this is a very partial list); however, it relies on the assumption that the classical action is itself supersymmetric. 
It is in this context that the work of Parisi and Sourlas~\cite{parisi_sourlas} provides a fresh perspective on how supersymmetry can be relevant for physical systems, by providing an additional way of understanding how the superpartners can appear and how supersymmetry can be realized and  broken. The idea is that the classical action need not be supersymmetric at all. Supersymmetry appears as the symmetry relating the fields, that appear in the classical action, to the fields that can resolve the fluctuations. In this context the supersymmetry of the classical action is independent of the supersymmetry that relates the fields of the classical action to the fields that ca, resolve the fluctuations.

The object of the present contribution is to provide a synthesis of trying to understand how it may shed complementary light. 

The plan of the paper is the following:

In section~\ref{Parisi_SourlasSUSY} we recall the idea of Parisi and Sourlas; in section~\ref{Anomalies} we discuss how anomalies appear in the simplest setting, that of toy models in 0-dimensional worldvolume; in section~\ref{Dim1wv} we show how tunneling in quantum mechanical models affects the anomalies and in section~\ref{Dim2wv} how two--dimensional models can avoid them. 
In \ref{Dim3beyondwv} we discuss how theories in higher dimensional worldvolumes might become susceptible.

The discussion is set within the framework of Wess--Zumino models, that provide a--limited, of course--framework for understanding the scalar sector of the Standard Theory; we comment on the challenges of extending this approach to gauge theories in the final section.

\section{Parisi--Sourlas supersymmetry}\label{Parisi_SourlasSUSY}
In 1982 Parisi and Sourlas made a remarkable observation~\cite{parisi_sourlas}: They observed  that the canonical partition function of a physical system, schematically written as
\begin{equation}
\label{Zcan}
Z = \int\,[\mathscr{D}\phi]\,e^{-S[\phi]}
\end{equation}
does not provide a complete description of the system along with the bath of fluctuations; one way to notice this is to realize that the canonical partition function is a function of the coupling constant(s). However the existence of the partition function, the fact that it is finite, implies that $S[\phi]$ can be expressed as 
$$
S[\phi] = \int\,dx\,\frac{1}{2}\left(\frac{\partial U}{\partial \phi(x)}\right)^2
$$
in order that $S[\phi]$ be bounded from below; it should, also, confine at infinity, in order to define a finite partition function.
Indeed, as was noticed by Parisi and Sourlas, we have the identity
\begin{equation}
\label{noiseid}
\int\,[\mathscr{D}\phi]\,
\,\left|\mathrm{det}\frac{\partial^2 U}{\partial \phi(x)\partial \phi(y)}\right|\,e^{-\int\,dx\,\frac{1}{2}\left(\frac{\partial U}{\partial \phi(x)}\right)^2}=1
\end{equation}
for an appropriate choice of units.

We may write 
$$
\left|\mathrm{det}\frac{\partial^2 U}{\partial \phi(x)\partial \phi(y)}\right|=
e^{-\mathrm{i}\theta_\mathrm{det}}\,\mathrm{det}\frac{\partial^2 U}{\partial \phi(x)\partial \phi(y)}
$$
and introduce the determinant in the exponent by using Grassmann variables
$$
\mathrm{det}\frac{\partial^2 U}{\partial x\partial y}=\int\,[\mathscr{D}\psi][\mathscr{D}\chi]\,e^{\int\,dx\,dy\,\chi(x)\frac{\partial^2U}{\partial\phi(x)\partial \phi(y)}\psi(y)}
$$
and check that the action
$$
S[x,\psi,\chi]=\int\,dx\,\frac{1}{2}\left(\frac{\partial U}{\partial \phi(x)}\right)^2-\int\,dx\,dy\,\chi(x)\frac{\partial^2U}{\partial \phi(x)\partial \phi(y)}\psi(y)
$$
is invariant under transformations that are defined by anticommuting parameters (Grassmann variables),$\zeta$ and $\eta,$ namely
$$
\begin{array}{l}
\displaystyle
\delta \phi(x) = \zeta\psi(x) + \chi(x)\eta\\
\displaystyle
\delta\chi(x) =-\eta\frac{\partial U}{\partial\phi(x)}\\
\displaystyle
\delta\psi(x) = -\zeta\frac{\partial U}{\partial\phi(x)} 
\end{array}
$$ 
These transformations realize ``worldpoint'' supersymmeytry, if $U(\phi)$ does not contain derivatives with respect to the worldvolume labels; if $U(\phi)$ does contain derivatives, then these will realize ``worldvolume'' or ``target space'' supersymmetry, depending on the properties of the operator $\partial U/\partial\phi(x).$

Furthermore, if $U(\phi)$ is not a quadratic function(al) of $\phi,$ these are non-linear transformations. They can be linearized by introducing (an) auxiliary field(s), $F(x),$ defined by 
\begin{equation}
\label{auxfield}
F(x) = \frac{\partial U}{\partial\phi(x)}
\end{equation}

We remark that we do need to introduce two, independent, Grassmann variables, in order to exponentiate the determinant of the operator $\partial^2U/\partial\phi(x)\partial\phi(y).$

The important point to abstract from the above summary of ref.~\cite{parisi_sourlas} is its generality. The only thing that matters is that the operator $\partial^2U/\partial\phi(x)\partial\phi(y)$ shouldn't have a non-zero measure of zeromodes.  Nowhere have we committed to the dimensionality of the worldvolume (labeled by $x,y$ and so on) nor about the target space (the number of components of the scalar or the anticommuting fields). What we have committed to is that $(\partial U/\partial\phi(x))^2$ be real; which means that, if $\phi(x)$ takes complex values--or if $\partial U/\partial\phi(x)$ takes complex values--then what is meant by $(\partial U/\partial\phi(x))^2$ is $|\partial U/\partial\phi(x)|^2;$ with corresponding complications for the second derivative. 

What has been misunderstood since the publication of ref.~\cite{parisi_sourlas} is that, insofar as the scalar, $\phi(x)$ is a ``physical'' field, so are its superpartners, $\psi(x)$ and $\chi(x)$ (and vice versa). Neither can be ghosts. Rather, either $\phi,\psi$ and $\chi$ all are ghosts (which is what happens with BRST transformations), or none are. 

In particular, if we want that the anticommuting fields, $\psi(x)$ and $\chi(x),$ represent target space fermions, then (and only then)  $\partial^2U/\partial\phi(x)\partial\phi(y)$ must be a Dirac operator. 

 We must appreciate that the compact notation used in ref.~\cite{parisi_sourlas} hides many subtle issues, that we shall attempt to elucidate in what follows. 
 
\section{Anomalies}\label{Anomalies}
The above considerations imply that $F(x)$ is a Gaussian field, with ultra--local 2--point function,
\begin{equation}
\label{FFvev}
\langle F(x)F(x')\rangle = 2\delta(x-x')
\end{equation}
and that the 1--point function vanishes
\begin{equation}
\label{Fvev}
\langle F(x)\rangle = 0
\end{equation}
while the higher order correlation functions are given by Wick's theorem. 

Indeed, eq.~(\ref{noiseid}), along with the definition of $F(x),$ (eq.~(\ref{auxfield})), can be understood as describing the change of variables in the partition function, from the field(s) $\phi(x)$ to the field(s) $F(x).$  

So there are two questions that can be raised: 
\begin{enumerate}
\item If we work with the partition function, as implied by eq.~(\ref{noiseid}) and compute the correlation functions of $F(x)=\partial U/\partial\phi(x),$ do they satisfy the identities~(\ref{FFvev}), (\ref{Fvev}) and the others, implied by Wick's theorem, or do they show anomalies?
\item Can the fluctuations produce the contribution $|\mathrm{det}\,\partial^2U/\partial\phi(x)\partial\phi(y)|$ or must it be included by hand?
\end{enumerate}
Now what is known is that questions about fluctuations probe the role of  the dimensionality of the worldvolume and of the target space. So we shall address these questions in terms of the dimensionality of the worldvolume and, for each case thereof, discuss how the properties of the target space affect the discussion.

The case of zero--dimensional worldvolume (for one--dimensional target space) was the subject of ref.~\cite{nicolis_zerkak}. There the first issue was addressed. The second issue was left open--but it is clear that the answer is No: Fluctuations cannot produce the contribution of $|\mathrm{det} \partial^2 U/\partial\phi(x)\partial\phi(y)|.$ This can be immediately checked for the case of integrals of one variable: If we attempt to compute the identities that ought to be satisfied by the noise fields, if the absolute value of the Jacobian is not explicitly included, we will find that they are not satisfied. 

A sample computation can be readily done for the case $dU/d\phi = m\phi + (\lambda\phi^2/2)=F(\phi).$ We can readily find that the answer is negative. And the reason isn't hard to understand: The absence of tunneling, since nothing can ``propagate''. This is what changes, when we have one--dimensional worldvolumes.
\section{One--dimensional worldvolume }\label{Dim1wv}
This is the case of the single particle, in a bath of fluctuations. In this case, 
\begin{equation}
\label{dUdphi1}
\frac{1}{2}\left(\frac{\partial U}{\partial\phi(x)}\right)^2=\frac{1}{2}\left(\frac{d\phi}{dx}\right)^2 + V(\phi(x))
\end{equation}
where $x$ is the Euclidian time. 

Now $V(\phi)$ must be bounded from below, therefore can be written as 
$$
V(\phi(x))=\frac{1}{2}\left(\frac{dW}{d\phi}\right)^2
$$
and we remark that 
$$
\frac{1}{2}\left(\frac{d\phi}{dx}\right)^2 + \frac{1}{2}\left(\frac{dW}{d\phi}\right)^2=\frac{1}{2}\left( \frac{d\phi}{dx} \pm \frac{dW}{d\phi}\right)^2 \mp\frac{dW}{dx}
$$
allowing the identification 
\begin{equation}
\label{Nicolaimap}
\frac{\partial U}{\partial \phi(x)} = \frac{d\phi}{dx} \pm \frac{dW}{d\phi}
\end{equation}
and 
\begin{equation}
\label{Jacobian}
\frac{\partial^2U}{\partial\phi(x)\partial\phi(y)}=\delta(x-y)\frac{d}{dx}\pm \frac{\partial^2 W}{\partial\phi(x)\partial\phi(y)}
\end{equation}
If $W(\phi(x))$ is an ultralocal function of $\phi(x),$ then 
$$
\frac{\partial^2 W}{\partial\phi(x)\partial\phi(y)}=\delta(x-y)\frac{\partial^2W}{\partial\phi(x)^2}
$$
and thus 
$$
\frac{\partial^2U}{\partial\phi(x)\partial\phi(y)}=\delta(x-y)\left\{\frac{d}{dx}\pm\frac{\partial^2W}{\partial\phi(x)^2}\right\}
$$
The upshot is that the auxiliary field, $F(x),$ is given by the expression
\begin{equation}
\label{auxfield1d}
F(x)=\frac{d\phi}{dx}\pm\frac{\partial W}{\partial\phi(x)}
\end{equation}
where we can choose the sign. 

We remark that, assuming periodic boundary conditions, we have that 
\begin{equation}
\label{vevF1}
\langle F(x)\rangle = \left\langle\frac{\partial W}{\partial\phi(x)}\right\rangle
\end{equation}
while the 2--point function is given by 
\begin{equation}
\label{vevF2}
\begin{array}{l}
\displaystyle
\langle F(x) F(x')\rangle=\left\langle\left(\frac{d\phi}{dx}\pm\frac{\partial W}{\partial\phi(x)}\right)
\left(\frac{d\phi}{dx'}\pm\frac{\partial W}{\partial\phi(x')}\right)\right\rangle
\end{array}
\end{equation}
It is possible to compute this correlation function by Monte Carlo simulations~\cite{Nicolis:2016osp,Nicolis:2014yka}, which provide evidence that no anomalies arise; there are lattice artifacts, that don't imply additional terms to the identities that are satisfied by the correlation functions of the noise fields, $F(x).$

What is an open question, that will be the subject of forthcoming work, is whether perturbation theory, about free fields, can provide an anomaly--free description.

\section{Two--dimensional worldvolume}\label{Dim2wv}
When the worldvolume is two--dimensional, further complications appear.  We would like to identify the term $(\partial U/\partial\phi(x))^2$ with the kinetic term, 
$$
\left(\frac{\partial\phi}{\partial x}\right)^2 + \left(\frac{\partial\phi}{\partial y}\right)^2
$$
and, also, ensure that the second derivative, $\partial^2U/\partial\phi(x)\partial\phi(y),$ has the properties expected of the two--dimensional Dirac operator, in order that, upon introducing its determinant in the action, using anticommuting fields, these can be identified as target space fermions. 

Now the experience with how Dirac derived his equation in the first place shows that the way to make this possible is to introduce two scalars, two noise fields and write $\partial U/\partial\phi(x)$ as 
\begin{equation}
\label{noiseid2d}
\begin{array}{l}
\displaystyle
F_1(\phi_1,\phi_2) = \partial_x\phi_2 + \partial_y\phi_1 + \frac{\partial W}{\partial\phi_1}\\
\displaystyle
F_2(\phi_1,\phi_2) = \partial_x\phi_1 - \partial_y\phi_2 + \frac{\partial W}{\partial\phi_2}
\end{array}
\end{equation}
It is then straightforward to deduce that 
$$
\frac{1}{2}\left(F_1^2 + F_2^2\right) = 
\frac{1}{2}\left(
\left(\frac{\partial\phi_1}{\partial x}\right)^2 + \left(\frac{\partial\phi_1}{\partial y}\right)^2 + \left(\frac{\partial\phi_2}{\partial x}\right)^2 + \left(\frac{\partial\phi_2}{\partial y}\right)^2\right) +  \frac{1}{2}\left(  \left(\frac{\partial W}{\partial\phi_1}\right)^2 + \left(\frac{\partial W}{\partial\phi_2}\right)^2 \right) + \mathrm{total\,derivatives} + \mathrm{crossterms}
$$
The crossterms are 
\begin{equation}
\label{crossterms2d}
\mathrm{crossterms} = \left(\partial_x\phi_2 + \partial_y\phi_1\right)\frac{\partial W}{\partial\phi_1} + \left(\partial_x\phi_1 - \partial_y\phi_2\right)\frac{\partial W}{\partial\phi_2}
\end{equation}
These terms explicitly break SO(2) rotation invariance, so, if they aren't total derivatives, as well, they can lead to anomalies. 
Parisi and Sourlas argue that, if the scalars satisfy the Cauchy--Riemann identities, then the crossterms vanish, whatever the superpotential. The problem with that argument is that it implies that the scalars are (a) massless and (b) satisfy the equations of motion for free, massless, fields. These would preclude the presence of a superpotential, then, at all. 

However, if we take the example they present, namely the choice 
\begin{equation}
\label{WZ2d}
\begin{array}{l}
\displaystyle
\frac{\partial W}{\partial\phi_1} = g(\phi_1^2-\phi_2^2)\\
\displaystyle
\frac{\partial W}{\partial\phi_2} = 2sg\phi_1\phi_2
\end{array}
\end{equation}
where $s=\pm1,$ one finds that the crossterms take the form
\begin{equation}
\label{crossterms2dWZ}
\mathrm{crossterms} = \partial_x\left(\phi_1^2\phi_2-\frac{\phi_2^3}{3}  \right) + \partial_y\left(\frac{\phi_1^3}{3}-\phi_1\phi_2^2 \right)+
 + 2g\phi_1\phi_2(s-1)\left(\partial_x\phi_1-\partial_y\phi_2\right)
\end{equation}
which implies that the crossterms are total derivatives iff $s=1.$ However, the superpotential is a hiolomorphic function of the scalars, i.e. it satisfies the Laplace equation,
\begin{equation}
\label{LaplaceWZ}
\frac{\partial^2W}{\partial\phi_1^2}+\frac{\partial^2W}{\partial\phi_2^2}=2g\phi_1+2gs\phi_1=2g(s+1)\phi_1
\end{equation}
iff $s=-1.$ 

Therefore the choice is between a non-holomorphic superpotential, that is, however, consistent with SO(2) coordinate invariance and a holomorphic superpotential, that breaks SO(2) coordinate invariance. Now it ought to be clear that SO(2) coordinate invariance is more 
significant (among other reasons it is the Euclidian avatar of Lorentz invariance). 

Monte Carlo simulations for the choice $s=1$ provide evidence that the noise fields do satisfy Wick's theorem to numerical precision, without any anomalies~\cite{Nicolis:2017lqk}. What remains to be elucidated is, what happens if we take $s=-1.$

\section{Beyond two spacetime dimensions }\label{Dim3beyondwv}
Parisi and Sourlas tried to go beyond two spacetime dimensions and seemed to hit an obstacle. However, the obstacle was, on the one hand, the assumption that the superpotential must be a holomorphic function of the fields and, on the other hand, the fact that, in Euclidian signature, if the worldvolume dimension $D$ was not equal to 2 mod 8, the Dirac matrices could not have only real entries. 
As we saw above--and as is well-known--for the Wess--Zumino model, holomorphic anomalies are expected to appear. So what remains is to understand how to deal with the fact that Dirac matrices in Euclidian signature, in spacetime dimensions $D=3$ and $D=4$ that are particularly relevant for applications to condensed matter and particle physics, have imaginary entries. 

A first conceptual issue has to do with the concern that Dirac matrices are assumed to carry  spinor indices, however they appear, also, in the kinetic terms for the scalars. However, if we examine the expressions for the noise fields, $F_1$ and $F_2,$~(\ref{noiseid2d}), we realize that the RHS of these expressions are perfectly reasonable expressions for scalars; what is interesting is that they can, indeed, be expressed more compactly as
\begin{equation}
\label{noiseid2dmatrix}
F_I(\phi_1,\phi_2) = \sigma_{IJ}^\mu\partial^\nu\delta_{\mu\nu}\phi_J +\frac{\partial W}{\partial\phi_I}
\end{equation}
where $\sigma_1 = \sigma_x$ and $\sigma_2=\sigma_z$ are the Pauli matrices. This, however, doesn't imply, by itself, anything about the spacetime transformations of the fields $\phi_I.$ They can be--and in these expressions are--scalars. 

On the other hand, we can understand these same equations as defining a change of variables in the partition function for the scalars, from the scalars $\phi_1$ and $\phi_2$ to the noise fields, $F_1$ and $F_2.$ This change of variables has as Jacobian the determinant of the differential operator
\begin{equation}
\label{Jacobian}
\mathcal{J}_{IJ} = \sigma_{IJ}^\mu\partial^\nu\delta_{\mu\nu} + \frac{\partial^2W}{\partial\phi_I\partial\phi_J}
\end{equation} 
and when we introduce this operator in the action using anticommuting variables, these do have the properties of Euclidian spinors. 
The relations~(\ref{noiseid2d}) and (\ref{noiseid2dmatrix}) are now known as ``Nicolai maps'': They provide a way of describing the effects of fermions in terms of their superpartners. 

They were introduced by Nicolai~\cite{Nicolai:1980jc,Nicolai:1980js} for simplifying the calculations for supersymmetric theories; it was the idea of Parisi and Sourlas that these maps apply to any theory, even if the classical action is not manifestly supersymmetric.

Furthermore, since the matrix $\partial^2W/\partial\phi_I\partial\phi_J$ is not, necessarily, symmetric (if the superpotential isn't a holomorphic function of the scalars) this can have interesting consequences (some of which were hinted in~\cite{parisi_sourlas} but remain to be spelled out). 

So the way to go around the obstacle is quite straightforward: It is necessary to double the degrees of freedom and deal with complex valued fields~\cite{Nicolis:2021buh}. We saw an example of this in two spacetime dimensions, when we needed two scalars, that can be identified as the real and imaginary parts of one complex doublet. 

In $D=3,$ where one, at least, of the Pauli matrices, has imaginary entries and we do need all three, we, therefore, require, at least, three complex doublets (six real scalars); and, in $D=4,$ relevant for particle physics, we need, at least, four complex doublets (eight real scalars). These considerations lead to predictions about the fields that are required at a minimum, in order that fluctuations may be resolved in terms of particles and the fields that create them. The possible anomalies are to be probed in the identities of the correlation functions of the noise fields (that define the corresponding Nicolai maps). Such anomalies will show up, if the crossterms aren't total derivatives, similar to what happens in two spacetime dimensions.

\section{Conclusions and outlook}\label{concl}
The upshot of our analysis is that supersymmetry can appear in two ways~\cite{Nicolis:2023mre,Nicolis:2025txm}: Either in the classical action--where the superpartners do not resolve the fluctuations--or when they do. In the former case there is considerable freedom and supersymmetry is discretionary; in the latter things are much more constrained and supersymmetry appears as inevitable.

The obstacles that have held up progress can be evaded. Regarding further immediate work, it is worth mentioning that 
an interesting point in this context is what happens if the {\em target space}  isn't one--dimensional; if the particle can explore a higher dimensional target space. In particular, if the target space is three--dimensional, the classical equations of motion may describe deterministic chaos~\cite{Nicolis:2019yem,Ovchinnikov:2025rvd,ovchinnikov2016introduction} and it is tempting to try and describe the chaotic fluctuations of nonintegrable, classical systems, in terms of the superpartners of the position, in the presence of external fluxes, that are self--consistently determined. This study needs to be done properly. 

Another direction is that of integrability, where the relation to the studies carried out by the Dubna group~(cf. e.g.\cite{Ivanov:2025ahr,Ivanov:2019gxo}) deserves to be further explored. 

However it can be claimed that the construction of the Nicolai map(s) for the Wess--Zumino class of models is now understood and concrete computations can be done. Furthermore, that the Wess--Zumino models, with extended supersymmetry, provided the framework for describing the fluctuations of any scalar theory, respectively of Higgs--Yukawa models. In this context the idea pertaining to ``anti-unification''~\cite{Iliopoulos:1980zd} deserves another look. 

Of course the challenge now is the construction of the Nicolai map(s) for gauge theories. Theres been considerable work on this, especially recently~\cite{Lechtenfeld:2021yjb,Lechtenfeld:2021uvs} but conceptual issues remain to be resolved. 

{\bf Acknowledgements:} I would like to thank the organizers of the Corfu Workshop(s) for the opportunity to give a talk on this subject and benefit from the stimulating environment they have fostered. Discussions with M. Axenides, E. Floratos, J. Iliopoulos and A. Schwimmer are gratefully acknowledged.
\bibliographystyle{JHEP}
\bibliography{SUSY}

\providecommand{\href}[2]{#2}\begingroup\raggedright\begin{thebibliography}{10}

\bibitem{bzowski2020consistency}
A.~Bzowski, G.~Festuccia and V.~Procházka, \emph{{Consistency of
  supersymmetric 't Hooft anomalies}},
  \href{https://arxiv.org/abs/2011.09978}{{\ttfamily 2011.09978}}.

\bibitem{katsianis2020supersymmetry}
G.~Katsianis, I.~Papadimitriou, K.~Skenderis and M.~Taylor,
  \emph{{Supersymmetry anomaly in the superconformal Wess-Zumino model}},
  \href{https://arxiv.org/abs/2011.09506}{{\ttfamily 2011.09506}}.

\bibitem{Papadimitriou:2017kzw}
I.~Papadimitriou, \emph{{Supercurrent anomalies in 4d SCFTs}},
  \href{http://dx.doi.org/10.1007/JHEP07(2017)038}{\emph{JHEP} {\bfseries 07}
  (2017) 038}, [\href{https://arxiv.org/abs/1703.04299}{{\ttfamily
  1703.04299}}].

\bibitem{Kuzenko:2019vvi}
S.~M. Kuzenko, A.~Schwimmer and S.~Theisen, \emph{{Comments on Anomalies in
  Supersymmetric Theories}},
  \href{http://dx.doi.org/10.1088/1751-8121/ab64a8}{\emph{J. Phys. A}
  {\bfseries 53} (2020) 064003},
  [\href{https://arxiv.org/abs/1909.07084}{{\ttfamily 1909.07084}}].

\bibitem{parisi_sourlas}
G.~Parisi and N.~Sourlas, \emph{{Supersymmetric Field Theories and Stochastic
  Differential Equations}},
  \href{http://dx.doi.org/10.1016/0550-3213(82)90538-7}{\emph{Nucl. Phys.}
  {\bfseries B206} (1982) 321--332}.

\bibitem{nicolis_zerkak}
S.~Nicolis and A.~Zerkak, \emph{{Supersymmetric probability distributions}},
  \href{http://dx.doi.org/10.1088/1751-8113/46/28/285401}{\emph{J. Phys.}
  {\bfseries A46} (2013) 285401},
  [\href{https://arxiv.org/abs/1302.2361}{{\ttfamily 1302.2361}}].

\bibitem{Nicolis:2016osp}
S.~Nicolis, \emph{{How quantum mechanics probes superspace}},
  \href{http://dx.doi.org/10.1134/S1547477117020248}{\emph{Phys. Part. Nucl.
  Lett.} {\bfseries 14} (2017) 357},
  [\href{https://arxiv.org/abs/1606.08284}{{\ttfamily 1606.08284}}].

\bibitem{Nicolis:2014yka}
S.~Nicolis, \emph{{A particle in equilibrium with a bath realizes worldline
  supersymmetry}},  \href{https://arxiv.org/abs/1405.0820}{{\ttfamily
  1405.0820}}.

\bibitem{Nicolis:2017lqk}
S.~Nicolis, \emph{{Probing the holomorphic anomaly of the $D=2, \mathcal{N}=2$,
  Wess-Zumino model on the lattice}},
  \href{http://dx.doi.org/10.1134/S1063779618050313}{\emph{Phys. Part. Nucl.}
  {\bfseries 49} (2018) 899--903},
  [\href{https://arxiv.org/abs/1712.07045}{{\ttfamily 1712.07045}}].

\bibitem{Nicolai:1980jc}
H.~Nicolai, \emph{{Supersymmetry without anticommuting variables}},  in
  \emph{{Unification of the fundamental particle interactions. Proceedings,
  Europhysics study conference, Erice, Italy, March 17-24, 1980}}, p.~689,
  1980.

\bibitem{Nicolai:1980js}
H.~Nicolai, \emph{{Supersymmetry and Functional Integration Measures}},
  \href{http://dx.doi.org/10.1016/0550-3213(80)90460-5}{\emph{Nucl. Phys.}
  {\bfseries B176} (1980) 419--428}.

\bibitem{Nicolis:2021buh}
S.~Nicolis, \emph{{The hidden fluxes, that control the fluctuations of scalar
  fields}}, \href{http://dx.doi.org/10.1088/1742-6596/2105/1/012003}{\emph{J.
  Phys. Conf. Ser.} {\bfseries 2105} (2021) 012003},
  [\href{https://arxiv.org/abs/2107.03194}{{\ttfamily 2107.03194}}].

\bibitem{Nicolis:2023mre}
S.~Nicolis, \emph{{Noisy SUSY}},
  \href{http://dx.doi.org/10.22323/1.436.0108}{\emph{PoS} {\bfseries CORFU2022}
  (2023) 108}, [\href{https://arxiv.org/abs/2303.17875}{{\ttfamily
  2303.17875}}].

\bibitem{Nicolis:2025txm}
S.~Nicolis, \emph{{A tale of two SUSYs}},
  \href{http://dx.doi.org/10.22323/1.490.0133}{\emph{PoS} {\bfseries CORFU2024}
  (2025) 133}, [\href{https://arxiv.org/abs/2503.23575}{{\ttfamily
  2503.23575}}].

\bibitem{Nicolis:2019yem}
S.~Nicolis, \emph{{Supersymmetry and deterministic chaos}},
  \href{http://dx.doi.org/10.1134/S1547477120050295}{\emph{Phys. Part. Nucl.
  Lett.} {\bfseries 17} (2020) 671--674},
  [\href{https://arxiv.org/abs/1912.12925}{{\ttfamily 1912.12925}}].

\bibitem{Ovchinnikov:2025rvd}
I.~V. Ovchinnikov, \emph{{On the Topological Nature of the Butterfly Effect}},
  \href{https://arxiv.org/abs/2503.18124}{{\ttfamily 2503.18124}}.

\bibitem{ovchinnikov2016introduction}
I.~V. Ovchinnikov, \emph{Introduction to supersymmetric theory of stochastics},
  {\emph{Entropy} {\bfseries 18} (2016) 108}.

\bibitem{Ivanov:2025ahr}
E.~Ivanov, \emph{{{\ensuremath{\mathscr{N}}} = 2 Higher Spins by Harmonic
  Superspace Methods}}, {\emph{Bulg. J. Phys.} {\bfseries 52} (2025) 018--031},
  [\href{https://arxiv.org/abs/2512.00573}{{\ttfamily 2512.00573}}].

\bibitem{Ivanov:2019gxo}
E.~Ivanov, O.~Lechtenfeld and S.~Sidorov, \emph{{Deformed $N$= 8 Supersymmetric
  Mechanics}}, \href{http://dx.doi.org/10.3390/sym11020135}{\emph{Symmetry}
  {\bfseries 11} (2019) 135}.

\bibitem{Iliopoulos:1980zd}
J.~Iliopoulos, D.~V. Nanopoulos and T.~N. Tomaras, \emph{{Infrared stability or
  anti-grand unification}},
  \href{http://dx.doi.org/10.1016/0370-2693(80)90843-6}{\emph{Phys. Lett. B}
  {\bfseries 94} (1980) 141}.

\bibitem{Lechtenfeld:2021yjb}
O.~Lechtenfeld and M.~Rupprecht, \emph{{Construction method for the Nicolai map
  in supersymmetric Yang-Mills theories}},
  \href{https://arxiv.org/abs/2104.09654}{{\ttfamily 2104.09654}}.

\bibitem{Lechtenfeld:2021uvs}
O.~Lechtenfeld and M.~Rupprecht, \emph{{Universal form of the Nicolai map}},
  \href{https://arxiv.org/abs/2104.00012}{{\ttfamily 2104.00012}}.

\end{thebibliography}\endgroup

\end{document}